\documentclass[twocolumn]{aastex62}

\hypersetup{linkcolor=blue,citecolor=blue,filecolor=cyan,urlcolor=magenta}

\graphicspath{{./}{figures/}}

\accepted{November 19, 2018}
\submitjournal{ApJ Letters}

\shorttitle{Deep Learning Exoplanet Classification}
\shortauthors{2018 NASA FDL Exoplanet Team}

\usepackage{natbib}
\usepackage{amsmath}

\begin{document}

\title{Scientific Domain Knowledge Improves Exoplanet Transit Classification with Deep Learning}

\correspondingauthor{Megan Ansdell}
\email{ansdell@berkeley.edu}

\author{Megan Ansdell}
\affil{Center for Integrative Planetary Science, University of California at Berkeley, United States}

\author{Yani Ioannou}
\affil{Machine Intelligence Lab, University of Cambridge, United Kingdom}

\author{Hugh P. Osborn}
\affiliation{Aix Marseille Univ, CNRS, CNES, Laboratoire d'Astrophysique de Marseille, France}

\author{Michele Sasdelli}
\affiliation{Australian Institute for Machine Learning, University of Adelaide, Australia}

\collaboration{(2018 NASA Frontier Development Lab Exoplanet Team)}

\author{Jeffrey C. Smith}
\author{Douglas Caldwell}
\affil{NASA Ames Research Center, United States}
\affil{SETI Institute, United States}

\author{Jon M. Jenkins}
\affiliation{NASA Ames Research Center, United States}

\author{Chedy R\"{a}issi}
\affiliation{Institut national de recherche en informatique et en automatique, France}

\author{Daniel Angerhausen}
\affiliation{Center for Space and Habitability, University of Bern, Switzerland}

\collaboration{(2018 NASA Frontier Development Lab Exoplanet Mentors)}


\begin{abstract}

Space-based missions such as {\it Kepler}, and soon {\it TESS}, provide large datasets that must be analyzed efficiently and systematically. Recent work by \cite{Shallue2018} successfully used state-of-the-art deep learning models to automatically classify {\it Kepler} transit signals as either exoplanets or false positives; our application of their model yielded 95.8\% accuracy and 95.5\% average precision. Here we expand upon that work by including additional scientific domain knowledge into the network architecture and input representations to significantly increase overall model performance to 97.5\% accuracy and 98.0\% average precision. Notably, we achieve 15--20\% gains in recall for the lowest signal-to-noise transits that can correspond to rocky planets in the habitable zone. We input into the network centroid time-series information derived from {\it Kepler} data plus key stellar parameters taken from the {\it Kepler} DR25 and {\it Gaia} DR2 catalogues. We also implement data augmentation techniques to alleviate model over-fitting. These improvements allow us to drastically reduce the size of the model, while still maintaining improved performance; smaller models are better for generalization, for example from {\it Kepler} to {\it TESS} data. This work illustrates the importance of including expert domain knowledge in even state-of-the-art deep learning models when applying them to scientific research problems that seek to identify weak signals in noisy data. This classification tool will be especially useful for upcoming space-based photometry missions focused on finding small planets, such as {\it TESS} and PLATO.

\end{abstract}



\section{Introduction} 
\label{sec:intro}

The past twenty-five years have seen the flourishing of two contemporaneous yet disparate fields---that of exoplanets in astronomy, and that of deep learning in computer science; both have rapidly moved from predominantly theoretical to now largely data-driven regimes. For exoplanet science, this has been powered by the launch of wide-field, high-precision space telescopes designed to search for transiting exoplanets. These facilities---in particular NASA's {\it Kepler} Space Telescope \citep{Borucki2016}---have discovered more than 3,000 confirmed planets \cite[$>70$\% of the total known; e.g.,][]{Borucki2011,Mayo2018}, enabling exoplanet population statistics that are revolutionizing our understanding of the universe \cite[e.g.,][]{Dressing2015,Gaidos2016}. For deep learning \cite[see review in][]{Lecun2015}, large labelled datasets, increases in computational power, and modern techniques for training deep neural networks have brought about breakthroughs in computer vision, speech recognition, and natural language processing~\cite[e.g.,][]{Krizhevsky2012,Ioffe2015,He2016}.

These two fields now intersect in the detection and classification of transit-like signals in the large quantities of data from space-based observatories like {\it Kepler} and soon {\it TESS}. These data need to be efficiently and reliably vetted for false-positive signals, such as those caused by stellar eclipses and instrumental noise, which largely outnumber the true planet transit signals. In particular, when searching for low signal-to-noise transit signals (e.g., as for small rocky planets in the habitable zone), chance correlations of stochastic instrumental and stellar signals can mimic transiting planet signals, making it extremely difficult to identify real transits just above the noise floor of the data. Deep learning---a machine learning tool named for its use of computational layers---provides a means to tackle these challenges.

For these reasons, exoplanet transit classification was selected as a project for the 2018 NASA Frontier Development Lab\footnote{\url{http://frontierdevelopmentlab.org}} (FDL), an eight-week research incubator aimed at applying cutting-edge machine learning algorithms to challenges in the space sciences. NASA FDL teams consist of two machine learning experts and two space science researchers, with the aim of enabling more effective machine learning models with the help of scientific ``domain knowledge"---i.e., the information, insight, or intuition relevant to a specific problem that a domain expert can provide that may not be immediately obvious to others. 

The use of deep learning for automatically classifying candidate exoplanet transit signals has been previously explored by \cite{Shallue2018}, who developed a convolutional neural network trained on {\it Kepler} data (see also \citealt{Zucker2018} and \citealt{Pearson2018} for applications of neural networks to simulated transit data). They clearly demonstrated the successful application of deep learning to transit classification, however improvements could be made with the inclusion of additional scientific domain knowledge. In this Letter, we present results from the 2018 NASA FDL program that investigated these possibilities. All code and data used in this work is publicly available.\footnote{\url{http://gitlab.com/frontierdevelopmentlab/exoplanets}}

\begin{figure}
\includegraphics[width=8.5cm]{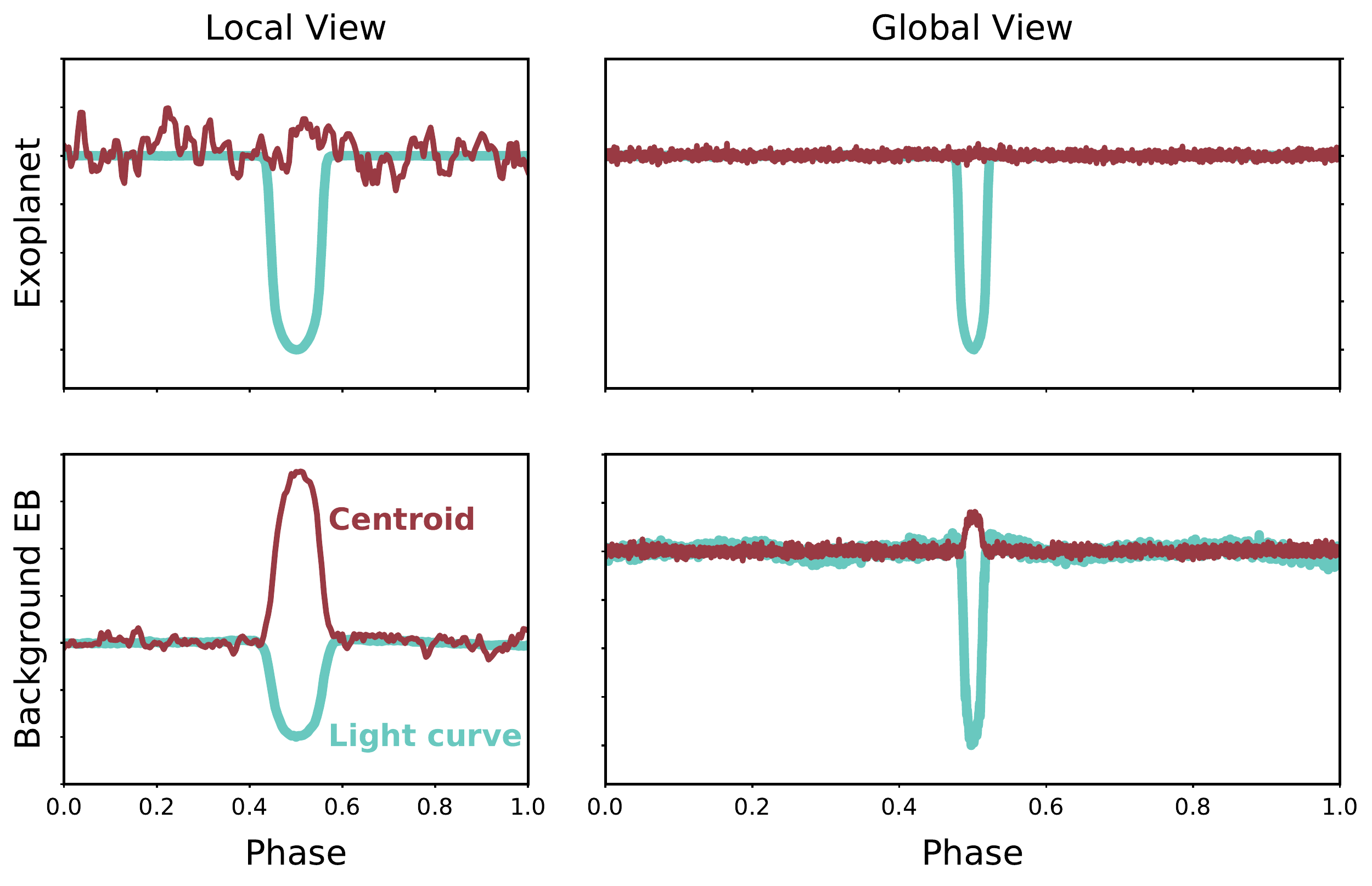}
\caption{\small The local (left) and global (right) views of the light curves (cyan) and centroids (maroon) for an example confirmed planet (top) and background eclipsing binary (bottom), illustrating how the centroid curves can be used to identify a common type of astrophysical false positive.}
\label{fig:views}
\end{figure}


\newpage
\section{Datasets} 
\label{sec:data}

\subsection{Flux Time Series (Light Curves)} 
\label{sec:flux}

In this work, we use the Q1--Q17 {\it Kepler} Data Release 24 (DR24) light curves from the Mikulski Archive for Space Telescopes.\footnote{\url{http://archive.stsci.edu/kepler}} These were produced by the {\it Kepler} Science Processing Pipeline \citep{Jenkins2010}, which starts by calibrating the time-series Target Pixel File (TPF) images, then performs fixed-aperture photometry, and removes systematic instrumental errors \cite[e.g.,][]{Smith2012, Stumpe2012}. Each light curve consists of integrated flux measurements spaced at 30-minute intervals spanning up to 4 years ($\simeq$70,000 points) and contains one or more ``threshold crossing events" (TCEs) identified by the {\it Kepler} pipeline. Each TCE is a potential exoplanet transit with a given period, epoch, and duration; however, most TCEs will be false-positive signals, sometimes caused by astrophysical phenomana such as eclipsing binaries (EBs) or background eclipsing binaries (BEBs), but also often by instrumental noise artifacts or other spurious events.

Following \cite{Shallue2018}, we perform additional processing of the light curves for each TCE. First, we flatten the light curve by iteratively fitting a basis spline (with the in-transit points of the TCE excluded to preserve the transit signal), then divide the light curve by the best-fit spline while linearly interpolating over the transit points (see \citealt{Shallue2018} for more details and also Figure~3 in \citealt{Vanderburg2014} for an illustration of this process). We implement a different spline-fitting routine than \cite{Shallue2018}, which results in 5$\times$ faster data processing times; namely, we use {\tt LSQUnivariateSpline} in {\tt SciPy} rather than {\tt bspline} in {\tt PyDL} (a library of Python replacements for IDL built-in functions). Second, we create ``global" and ``local" views of each phase-folded TCE following the description in \cite{Shallue2018}. Both views are scaled so that the continuum is at 0 and the maximum transit depth is at $-1$. The global view encapsulates the full view of the phase-folded light curve (e.g., including secondary transits of EBs) at the cost of long-period TCEs having poorly sampled transits. The local view, which depends on the transit duration, then provides a more detailed view of the primary transit shape. These two views are illustrated in Figure~\ref{fig:views} and used as inputs into the deep learning models implemented in this work (see Figure~\ref{fig-models}; Section~\ref{sec:models}).

\subsection{Centroid Time Series (Centroid Curves)} 
\label{sec:centroid}

We also use the time-series of the pixel position of the center of light (centroid) measured by the {\it Kepler} pipeline from the same TPF as the flux time series (Section~\ref{sec:flux}). Centroids provide information on the position of the source of the transit-like signal, which is particularly useful for identifying BEBs. This is because centroids will shift in the opposite direction of the BEB if both the BEB and target star are contained within the photometric aperture used to measure the flux. We use the flux-weighted (first moment) centroids rather than the pixel response function (PRF) centroids; although the PRF centroids are more robust against background noise, a significant number of sources do not have PRF centroid information, which complicates implementation of machine learning algorithms. 

We use the $x$ and $y$ pixel coordinates of the centroid to compute the absolute magnitude ($r=\sqrt{x^2 + y^2}$) of the centroid displacement from the TPF center. We then follow the same process as the light curves for smoothing, phase-folding, and translating into local and global views. However, rather than normalizing the centroid curve to the maximum transit depth, we subtract the median and divide by the standard deviation, where these values are calculated out-of-transit and across the entire training dataset (this standard practice is called ``normalization" in machine learning). Moreover, we normalize the standard deviation of the centroid curves by that of the light curves, which ensures that TCEs with no significant centroid shifts show flat lines with noise signal strengths similar to that of the light curves (and thus do not dominate the signal strengths). Example phase-folded light curves and associated centroid curves, for both global and local views, are given in Figure~\ref{fig:views} for a confirmed exoplanet and BEB.

\subsection{Stellar Parameters} 
\label{sec:stellar}

We use the updated {\it Kepler} DR25 catalogue of \citet{mathur2017revised} to obtain intrinsic stellar parameters. These parameters consist of stellar effective temperature ($T_{\rm eff}$), surface gravity ($\log g$), metallicity ([Fe/H]), radius ($R_\star$), mass ($M_\star$), and density ($\rho_\star$). These stellar parameters are normalized (Section~\ref{sec:centroid}): we subtract the median and divide by the standard deviation, where these values are calculated for each parameter across the entire training set, such that the distribution of each parameter has a median of 0.0 and standard deviation of 1.0. We note that we tested the inclusion of proper motions and parallaxes from the {\it Gaia} DR2 catalogue \citep{GaiaDR2}, but found no improvement in model performance.

\subsection{Labels} 
\label{sec:labels}

We use the same labels as \citet{Shallue2018}, which are taken from the {\it Kepler} DR24 TCE Table available on the NASA Exoplanet Archive.\footnote{\url{http://exoplanetarchive.ipac.caltech.edu/}} The {\bf av\_training\_set} column contains the labels used to train the {\tt Autovetter} \citep{mccauliff2015automatic} and primarily come from human-vetted KOIs assembled from multiple papers \citep[e.g.][]{Borucki2011,Batalha2013,Burke2014}. The {\bf av\_training\_set} column has four possible values: planet candidate (PC), astrophysical false positive (AFP), non-transiting phenomenon (NTP), and unknown (UNK). Following \citet{Shallue2018}, we ignore the UNK TCEs (4,630 entries) and then binarize the remaining labels as ``planet" (PC; 3,600 entries) or ``false positive" (AFP + NTP; 12,137 entries). We then randomly divide the TCEs into training (80\%), validation (10\%), and test (10\%) sets using the same random seed as \cite{Shallue2018} to preserve comparability. We use the validation set to tune hyperparameters and the test set for our final model performance results (Section~\ref{sec:results}).


\begin{figure*}
\begin{centering}
\includegraphics[width=15.5cm]{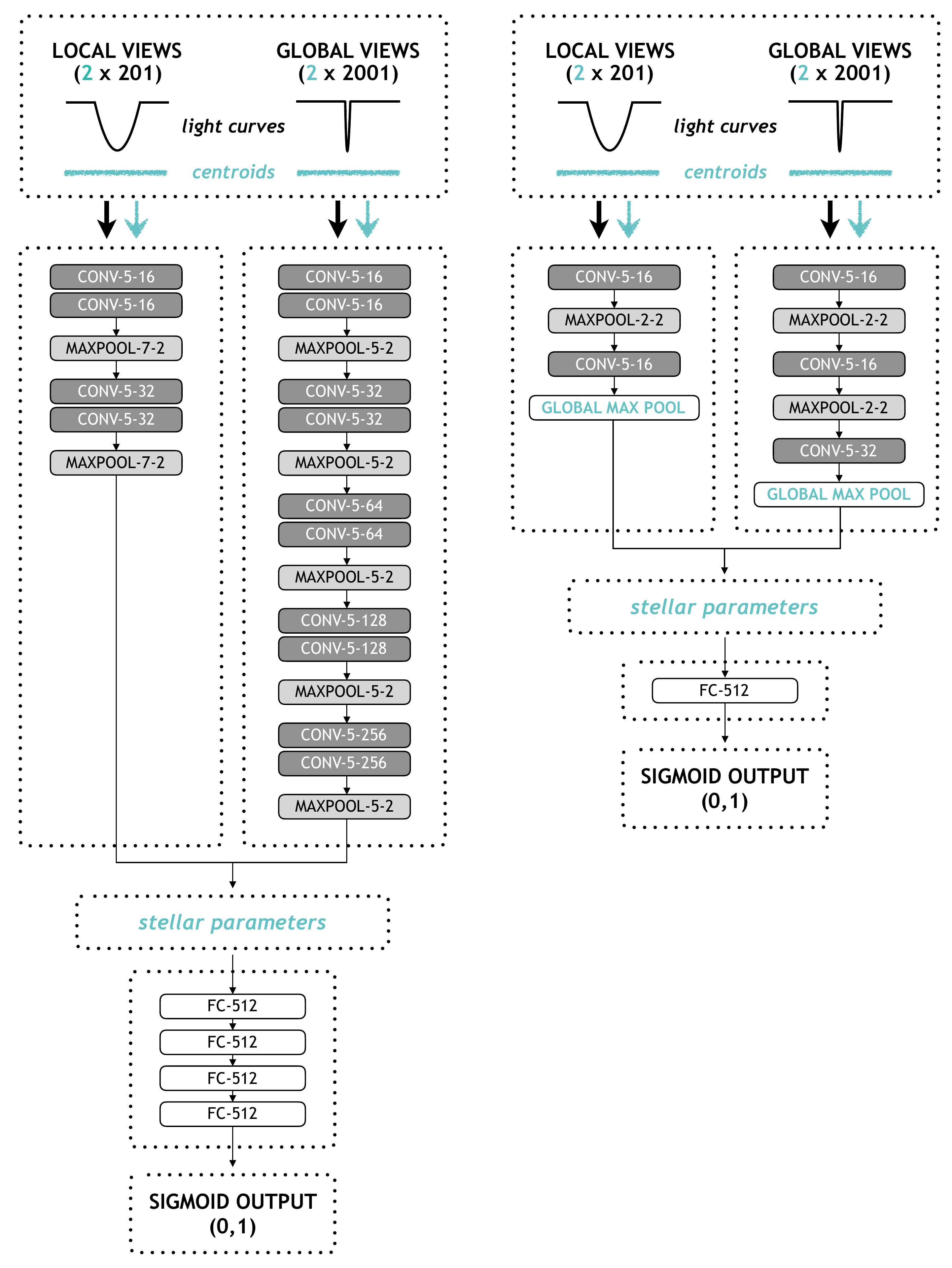}
\caption{\small The convolutional neural network architectures used in this work. {\bf Left:} {\tt Exonet}, where the additions over the baseline {\tt Astronet} model are shown in blue (Section~\ref{sec:Exonet}). The flattened outputs of the disjoint one-dimensional convolutional columns are concatenated with the stellar parameters, then fed into the fully connected layers ending in a sigmoid function. Following \citet{Shallue2018}, the convolutional layers are denoted as CONV-$\langle$kernel size$\rangle$-$\langle$number of feature maps$\rangle$, the max pooling layers are denoted as MAXPOOL-$\langle$window length$\rangle$-$\langle$stride length$\rangle$, and the fully connected layers are denoted as FC-$\langle$number of units$\rangle$. {\bf Right:} the significantly reduced {\tt Exonet-XS} model version described in Section~\ref{sec:small-Exonet}.}
\label{fig-models}
\end{centering}
\end{figure*}

\section{Machine Learning Models} 
\label{sec:models}

\subsection{{\tt Astronet}: Baseline Model}
\label{sec:astronet}

Here we briefly summarize {\tt Astronet}, the deep convolution neural network developed by \cite{Shallue2018} that we use as our baseline model. {\tt Astronet} is implemented in TensorFlow \citep{Abadi2016}, an open source software library for machine learning originally developed at Google Brain. As shown in Figure~\ref{fig-models}, the {\tt Astronet} model architecture has two disjoint one-dimensional convolutional columns (one for the global view and one for the local view) with max pooling, the results of which are concatenated and then fed into a series of fully connected layers ending in a sigmoid function that produces an output in the range (0,1) that loosely represents the likeliness of a given TCE being a true planet transit (1) or false positive (0).

For model training, {\tt Astronet} uses the Adam optimization algorithm \citep{Kingma2014} to minimize the cross-entropy error function. During training, data are augmented by applying time inversions to the input light curves with a 50\% chance. The Google-Vizier system \citep{46180} automatically tunes the hyperparameters of the input representations, model architecture, and training; consequently, the model is trained with a batch size of 64 for 50 epochs, and the Adam optimizer is implemented with a learning rate of $\alpha=10^{-5}$, exponential decay rates of $\beta_{1}=0.9$ and $\beta_{2}=0.999$, and $\epsilon=10^{-8}$ (see \citealt{Kingma2014} for details on these parameters). \cite{Shallue2018} use model ``ensembling" to average the results from 10 independently trained version of the same model with different random parameter initializations. Ensembling makes comparisons between different model architectures more robust because it averages over the stochastic differences in the individual models due to their different random parameter initializations. Moreover, ensembling improves model performance because the individual models can perform slightly better (or worse) in different regions of input space, in particular when the training set is small and thus prone to over-fitting.

As part of this work, we re-implemented {\tt Astronet} in PyTorch \citep{Paszke2017} in an effort to expand the user base to those unfamiliar with TensorFlow. Our {\tt Astronet} performance results given in Table~\ref{tab:results} are consistent with those reported in \cite{Shallue2018}; for example, we find an accuracy of 0.958 compared to their 0.960 value. We use the values in Table~\ref{tab:results} as our baseline for comparison in the rest of this work.

\subsection{{\tt Exonet}: Revised Model with Domain Knowledge}
\label{sec:Exonet}

Here we use scientific domain knowledge to add several features to our baseline {\tt Astronet} model architecture and input representations in an effort to increase model performance. This modified model, which we call {\tt Exonet}, is illustrated in Figure~\ref{fig-models} and compared to the baseline {\tt Astronet} model. The key modifications are described below. For model training, we retain the use of the Adam optimizer, cross-entropy loss function, batch size of 64, and learning rate of $\alpha=10^{-5}$ used by {\tt Astronet} (Section~\ref{sec:astronet}). 

{\bf Addition of centroid time-series}: we input our analogous global and local views of the centroid time-series data (Section~\ref{sec:centroid}) as second channels of the disjoint convolutional columns used for the light curves. The motivation behind this architecture is to help the model learn the connections between the shapes of the light curves and centroid curves, which can be useful for identifying false positives, in particular BEBs (Figure~\ref{fig:views}). We note that the addition of centroid information was suggested by \cite{Shallue2018} as a potential avenue for improvement.

{\bf Addition of stellar parameters}: we concatenate the stellar parameters (Section~\ref{sec:stellar}) to the flattened outputs of the convolutional layers directly before feeding them into the shared fully connected layers. We add this information because stellar parameters are likely correlated with classification, for example giant stars with large radii are far more likely to host stellar eclipses than planetary transits (which would be undetectable).

{\bf Augmentation of training data}: the baseline {\tt Astronet} model augments the data by randomly flipping the time axis of half the input light curves during training. We adopt this data augmentation technique, also applying the time-axis flip to the associated centroid curves. Because the training set is still quite small compared to typical machine learning problems, and because we found that {\tt Astronet} suffers from model over-fitting, we apply an additional data augmentation technique during training to mimic measurement uncertainties in the flux measurements. Namely, we add random Gaussian noise to each input light curve, where the standard deviation is randomly chosen from a uniform distribution between 0 and 1.

\subsection{{\tt Exonet-XS}: Reduced Model Size}
\label{sec:small-Exonet}

With the additions described in Section~\ref{sec:Exonet}, we find that we can drastically reduce the size of the model architecture while still maintaining improved performance. This reduced model architecture, which we call {\tt Exonet-XS}, is illustrated in Figure~\ref{fig-models}. We reduce the number of convolutional layers from 4 to 2 for the local column and from 10 to 3 for the global column. We also introduce global max pooling at the output of each convolutional column, as global max/average pooling has been shown to reduce the number of model parameters while increasing generalization \citep{Lin2013,Ioannou2016}, and is used in most state-of-the-art models for ImageNet \cite[e.g.,][]{He2016}. {\tt Exonet-XS} has a model size $\simeq$0.07\% that of the full {\tt Astronet} model with $\simeq$5$\times10^{-4}$ fewer trainable parameters. Smaller models are often preferred, as they generalize better (or over-fit less) \citep{Hastie2001}. Thus {\tt Exonet-XS} should perform better when re-trained on other datasets, for example those expected from {\it TESS}.
\\

\begin{deluxetable}{lcc}
\tablecaption{Ensembled Results on Test Set \label{tab:results}}
\tablecolumns{3}
\tablewidth{0pt}
\tablehead{
\colhead{Model} &
\colhead{Accuracy} &
\colhead{Avg. Precision}
}
\startdata
{\tt Astronet} & 0.958 & 0.955  \\
{\tt Exonet} & 0.975 & 0.980  \\
\hline
{\tt Astronet-XS} & 0.953 & 0.936  \\
{\tt Exonet-XS} & 0.966 & 0.963 
\enddata
\end{deluxetable}


\section{Results}
\label{sec:results}

To assess model performance, we use three key metrics: accuracy, average precision, and precision-recall curves. Accuracy is the fraction of correct classifications by the model, for both planets and false-positives, at a given threshold for deciding when the model output in the range (0,1) becomes a positive class; we use a threshold of 0.5 for our accuracy calculations. Precision is the fraction of transits classified as planets that are true planets, while recall is the fraction of true planets recovered by the model; these can then be plotted on a precision-recall curve, which shows the trade-off between precision and recall for different thresholds. The average precision summarizes the precision-recall curve as the weighted mean of precisions achieved at each threshold.

Table~\ref{tab:results} gives the ensembled results on the test set, showing the 1.7\% increase in accuracy and 2.5\% increase in average precision of {\tt Exonet} over {\tt Astronet}. For the reduced {\tt Exonet-XS} model, we see a 0.8\% increase in accuracy and 0.8\% increase in average precision. We also show results for the reduced model without the additions described in Section~\ref{sec:Exonet} ({\tt Astronet-XS)}, to illustrate that the gains in performance are still due to the inclusion of new scientific domain knowledge, rather than the change in model architecture.

\begin{figure}
\includegraphics[width=8.3cm]{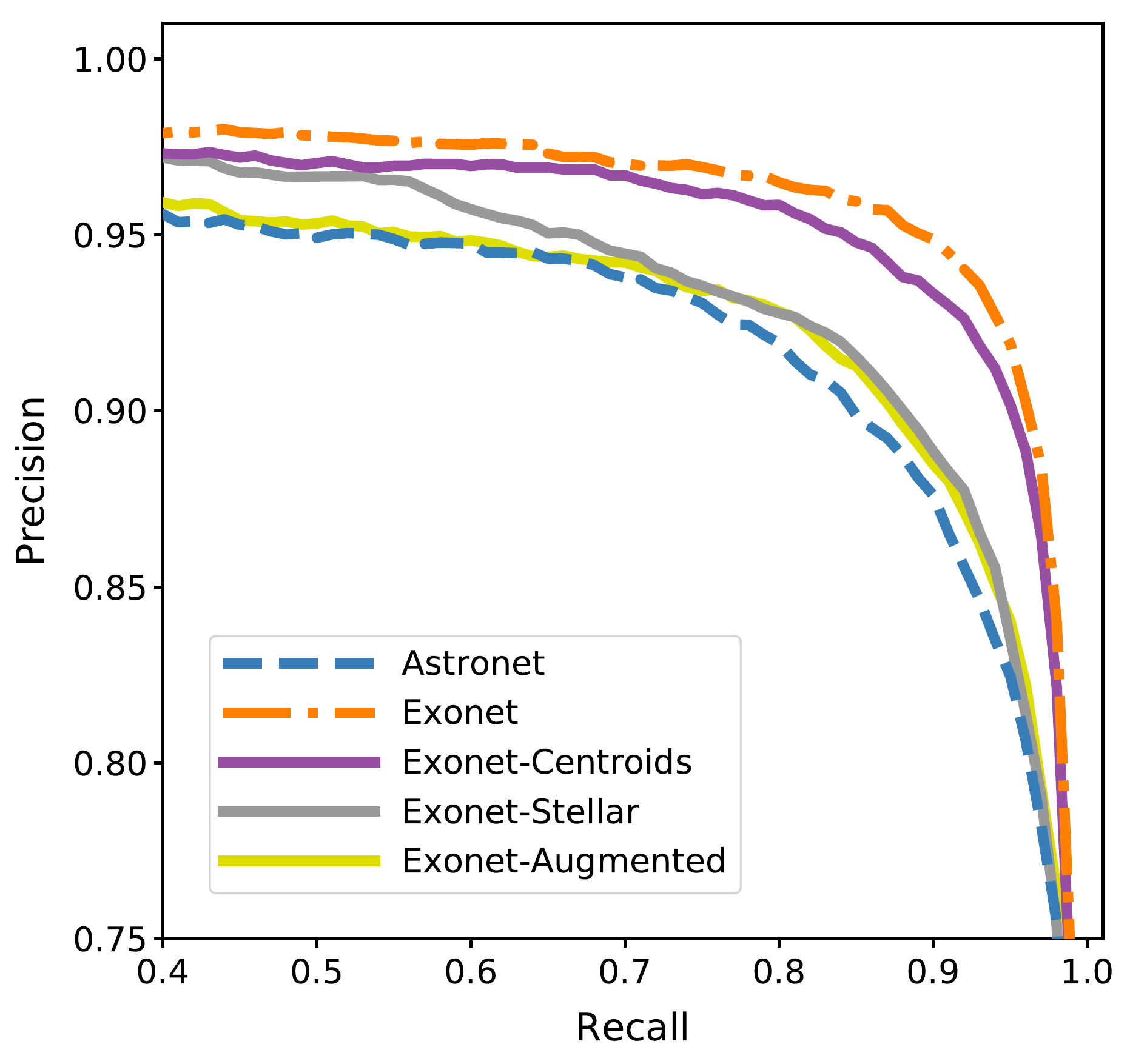}
\caption{\small Precision-recall curve of {\tt Astronet} (Section~\ref{sec:astronet}) compared to those of {\tt Exonet} (Section~\ref{sec:Exonet}) with different additions of scientific domain knowledge to show the individual contributions to increases in model performance. {\tt Exonet-Centroids} is just the addition of the centroid curves, {\tt Exonet-Stellar} is just the addition of the stellar parameters, and {\tt Exonet-Augmented} is just the addition of our supplementary data augmentations. {\tt Exonet} then combines all of these improvements into a single model.}
\label{fig-compare}
\end{figure}

Figure~\ref{fig-compare} shows precision-recall curves with each component of scientific domain knowledge added individually to illustrate their separate contributions to improving model performance. For this, we use $k$-fold cross-validation on the combined training and validation sets. Cross validation is a method of evaluating model generalization performance that is more robust than using training and test sets; in $k$-folds cross-validation, the data is instead split repeatedly into $k$ parts of equal size and multiple models are trained (here we use $k=5$, which is typical). Figure~\ref{fig-compare} shows that the addition of the centroid time series provides the biggest gain in model performance, while stellar parameters also make a significant impact. Data augmentation does not greatly increase model performance by itself, rather the main benefit is to alleviate model over-fitting.

\begin{figure}
\includegraphics[width=8cm]{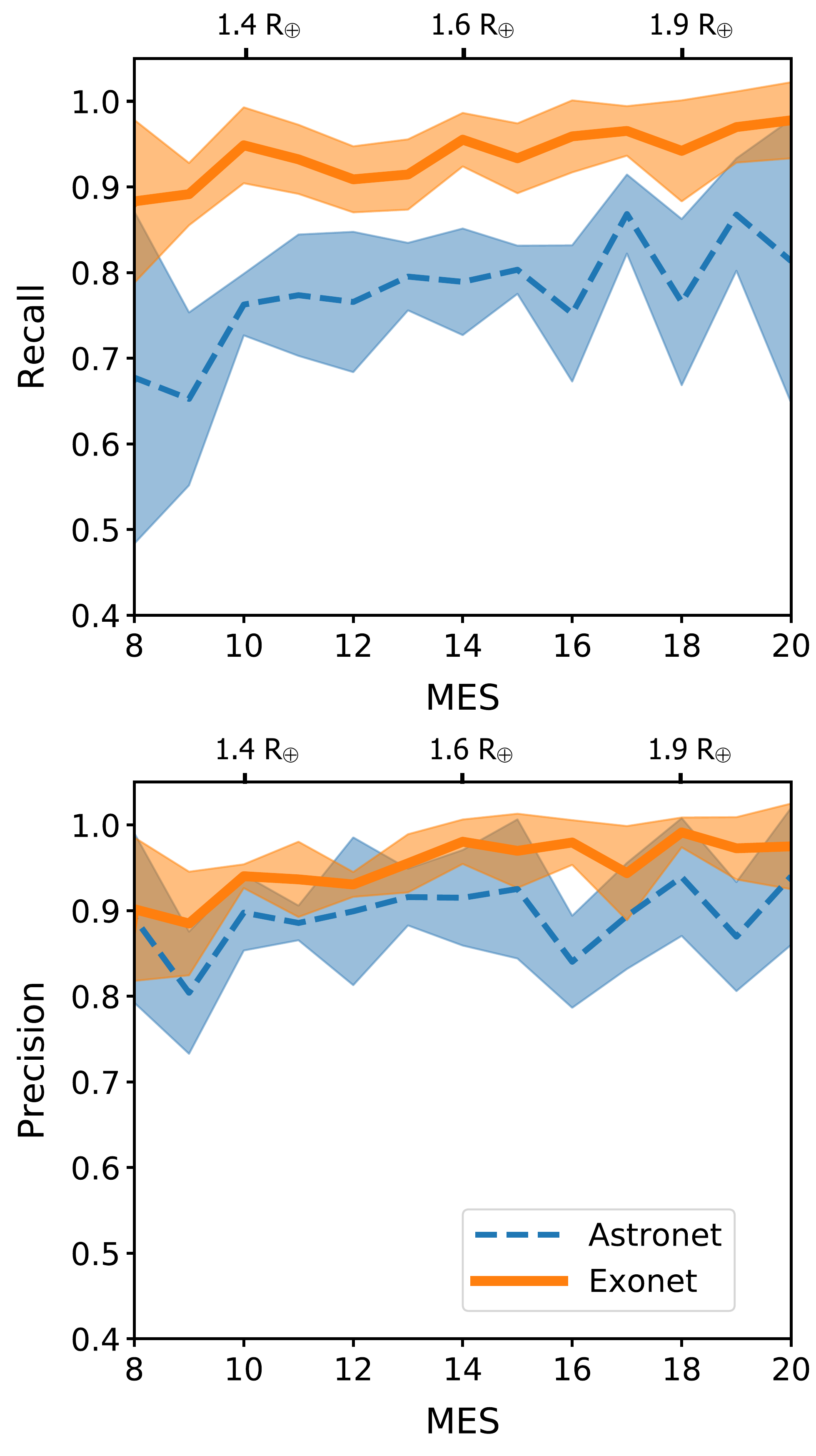}
\caption{\small Recall (top) and precision (bottom) as a function of MES, which is a measure of the signal-to-noise of candidate transits. The {\it Kepler} pipeline only reports MES$>$7.1 and the cross-validation $k$-folds sometimes did not contain any planets with MES$<8$, thus here we only plot MES$\ge$8. The solid lines are the averages of the cross-validation $k$-fold results, while the shaded regions show their standard deviations. We use a threshold of 0.7 to calculate the individual recall and precision values. The top axes show the median planet radius for confirmed/candidate KOIs in three MES bins, illustrating that the gains in performance by {\tt Exonet} can be most significant for Earth-sized planets.}
\label{fig-mes}
\end{figure}

Figure~\ref{fig-mes} then shows the precision and recall as a function of a measure of the signal-to-noise of the candidate transits---the so-called ``multiple event statistic" \cite[MES;][]{Jenkins2002} that the {\it Kepler} pipeline reports with each TCE. Notably, {\tt Exonet} shows 15--20\% increases in recall for low-MES transits that often correspond to Earth-sized planets, some of which are in the habitable zone. The scatter in model performance at a given MES value is also noticeably smaller for {\tt Exonet}, reflecting the robustness of the model. Note we do not perform hyperparameter optimization for {\tt Exonet} and {\tt Exonet-XS}, thus the model results could still be improved.

\section{Conclusions} 
\label{sec:summary}

We expanded upon the method of \citet{Shallue2018} for applying deep learning to automatically classify {\it Kepler} candidate transit events with the addition of scientific domain knowledge. We used as our baseline model their convolutional neural network architecture, {\tt Astronet}, which inputs ``global" and ``local" views of the phase-folded light curves through disjoint one-dimensional convolutional columns followed by shared fully connected layers that output a number in the range (0,1) that approximates the likeliness of a transit being a planet (1) or a false positive (0). 

For our modified model, which we call {\tt Exonet}, we created analogous global and local views of the centroid time-series data and input them as second channels of the disjoint convolutional columns, mainly to help identify BEBs. We then also concatenated key stellar parameters to the flattened outputs of the convolutional layers before feeding them into the shared fully connected layers; this helped to identify other types of false positives, such as giant star eclipsing binaries. Because we found that {\tt Astronet} was prone to over-fitting, we also implemented data augmentation during training. {\tt Astronet} already performed random time-axis reflections on the light curves, which we adopted and also applied to the corresponding centroid time series. Furthermore, we added random Gaussian noise to the input light curves to simulate uncertainties on the flux measurements.

These additions of scientific domain knowledge to the model architecture and input representations significantly improved the model accuracy and average precision by $\simeq$2.0--2.5\%, which is notable given the already impressive performance of {\tt Astronet} prior to this work. Moreover, we showed that these gains in model performance are disproportionately high for low signal-to-noise transits that can represent the most interesting cases of rocky planets in the habitable zone. This demonstrates the importance of including domain knowledge in even state-of-the-art machine learning models when applying them to scientific research problems that seek to identify weak signals in noisy data. This classification tool will be especially useful for upcoming space-based photometry missions focused on finding small planets, such as {\it TESS} \citep{Ricker2014} and PLATO \citep{Rauer2014}. A forthcoming paper will document the application of our deep learning model to simulated {\it TESS} data (Osborn et al., in prep).

\software{Astropy \citep{AstroPy2013,AstroPy2018}, PyTorch \citep{Paszke2017}, Astronet \citep{Shallue2018}, Jupyter \citep{Kluyver:2016aa}, SciPy \citep{SciPy2001}, TensorFlow \citep{Abadi2016}, Matplotlib \citep{Hunter2007}.}

\acknowledgments{We gratefully thank Google Cloud for providing the compute and storage resources critical to completing this work. MA also gratefully acknowledges the NVIDIA Corporation for donating the Quadro P6000 GPU used for this research. We thank Adam Lesnikowski, Noa Kel, Hamed Valizadegan, Yarin Gal, and Richard Galvez for their helpful discussions. This paper includes data from the {\it Kepler} mission; funding for {\it Kepler} is provided by the NASA Science Mission directorate. The data used in this paper were obtained from the Mikulski Archive for Space Telescopes (MAST). STScI is operated by the Association of Universities for Research in Astronomy, Inc., under NASA contract NAS5-26555. MA also acknowledges support from NSF grant AST-1518332 and NASA grants NNX15AC89G and NNX15AD95G/NEXSS. HPO acknowledges support from Centre National d'Etudes Spatiales (CNES) grant 131425-PLATO.}

\newpage


\bibliography{bib.bib}

\end{document}